\documentclass{article}
\usepackage{spconf,amsmath,graphicx}
\usepackage{amssymb}
\usepackage{amsmath}
\usepackage{listings}
\usepackage{hyperref}
\usepackage{subfig}
\usepackage{caption}

\title{IPP-Net: A Generalizable Deep Neural Network Model for Indoor Pathloss Radio Map Prediction}

\name{Bin Feng$^{1,2,*}$, Meng Zheng$^{1,*}$, Wei Liang$^{1}$, Lei Zhang$^{1,2}$ \thanks{This work was supported by the LRTP (XLYC2203148).}
}
\address{$^{1}$Shenyang Institute of Automation, Chinese Academy of Sciences\\
$^{2}$University of Chinese Academy of Sciences\\
$*$ These authors contributed equally to this work. }

\begin{document}
%
\maketitle
\begin{abstract}
\vspace{-0.1cm}
In this paper, we propose a generalizable deep neural network model for indoor pathloss radio map prediction (termed as IPP-Net). IPP-Net is based on a UNet architecture and learned from both large-scale ray tracing simulation data and a modified 3GPP indoor hotspot model. The performance of IPP-Net is evaluated in \textit{the First Indoor Pathloss Radio Map Prediction Challenge in ICASSP 2025}. The evaluation results show that IPP-Net achieves a weighted root mean square error of 9.501 dB on three competition tasks and obtains the second overall ranking.
\end{abstract}
\begin{keywords}
Radio map prediction, deep neural network, channel model, indoor pathloss
\end{keywords}
\vspace{-0.3cm}
\section{Introduction}
\label{sec:intro}
\vspace{-0.3cm}
The design of wireless communication systems strongly relies on accurate information on radio propagation channels. Pathloss radio map, as a vivid description to radio environments, plays a vital role in the planning and optimization of wireless communication systems. Conventional Radio Map Prediction (RMP) methods are based on either low-accuracy empirical channel models or computation-intensive ray tracing simulations. Recently, Deep Neural Network (DNN)-based RMP methods \cite{9354041}-\cite{9771088} combining accuracy with computational efficiency have been emerging as a promising alternative to conventional ones. However, existing works are designed mainly for outdoor radio environments. Unlike outdoor radio environments which are dominated by reflected field components, refractive signal energy passing through obstacles plays a more significant role in indoor radio environments. Thus, it is necessary to develop RMP methods tailored for indoor radio environments.

In this paper, we propose a generalizable DNN model for indoor pathloss RMP (IPP-Net) to address \textit{the First Indoor Pathloss Radio Map Prediction Challenge in ICASSP 2025}. We design a DNN model that adopts a UNet architecture, with indoor geometries annotated with reflectance and transmittance values, physical distance, a modified 3GPP Indoor Hotspot (InH) \cite{3gpp_tr38_901_2024} model and frequency bands as inputs of IPP-Net. A curriculum learning strategy that progressively increases the complexity of the learning tasks is employed to train IPP-Net. The evaluation results show that IPP-Net achieves a weighted Root Mean Square Error (RMSE) of 9.501 dB on three competition tasks and obtains the second overall ranking \cite{rmapChallenge2025}.

\textbf{Task: Indoor Pathloss Prediction.} The challenge provides three tasks, aiming to test the generalizability of submitted methods in unseen geometries, frequencies, and antenna radiation patterns.

\textbf{Dataset: Indoor Radio Map Dataset.} The challenge dataset \cite{c0ec-cw74-24} comprises pathloss radio maps generated by ray tracing in 25 indoor scenarios, 3 frequency bands, and 5 antenna radiation patterns. Each  input of the dataset is an $H \times W$ RGB image, where the first two channels indicate reflectance and transmittance (in dB), respectively. The third channel represents the physical distance between the transmitter and each pixel of the image. The target output of the dataset is an $H \times W$ grayscale image whose pixels denote pathloss values.

\vspace{-0.3cm}
\section{IPP-Net}
\label{sec:method}
\vspace{-0.3cm}
\subsection{Network architecture}
\vspace{-0.3cm}
Inspired by the success of UNet on pathloss prediction \cite{9354041}-\cite{9771088}, we design IPP-Net based on a UNet architecture that consists of an initial layer, five encoding/decoding layers, a bottleneck layer, skip connections and an output layer. The initial layer prepares the five-channel inputs. The encoding/decoding layers and the bottleneck layer share five consecutive dilated convolution layers and a standard convolution layer. While the encoding layers use max pooling for down-sampling, the decoding layers adopt transposed convolution for up-sampling. The number of filters in encoding layers starts at 32, doubling with each layer, while in decoding layers halves from 512. Skip connections are employed to concatenate feature maps from the encoding layers to the corresponding decoding layers. The output layer is a $1\times1$ convolution layer which produces grayscale radio maps.

\vspace{-0.3cm}
\subsection{Network inputs}
\vspace{-0.3cm}
The inputs to IPP-Net consist of five channels. The first three ones jointly characterize RGB images from the challenge dataset. The fourth channel (termed as model channel) is a pathloss radio map generated by using a modified 3GPP InH model. The fifth channel is a frequency channel that annotates the frequency band of the target output radio map.

In order to boost the learning efficiency and accuracy, we modify the 3GPP InH model by introducing a Non-Line-Of-Sight (NLOS) level matrix $L=\{L_{i,j}\} \in \mathbb{Z}_+^{H \times W}$ and a weight matrix $\Delta =\{\Delta_{i,j}\}\in \mathbb{R}^{H \times W}$. $L_{i,j}$ denotes the NLOS level at the grid $(i,j)$ and can be explained as follows: $L_{i,j} = 0$, indicating that a direct link exists between the grid $(i,j)$ and the transmitter. $L_{i,j}>0$, signifying that obstructions (e.g., walls or doors) stand between the grid $(i,j)$ and the transmitter. Based on the Bresenham’s line algorithm, we calculate $L_{i,j}$ by counting the number of obstructions between the grid $(i,j)$ and the transmitter. On the other hand, $\Delta_{i,j}$ is the weight of $L_{i,j}$ and can be learned with the UNet weights jointly. 

Finally, the model channel $M = \{M_{i,j}\} \in \mathbb{R}_+^{H \times W}$can be given as follows 

\vspace{-0.2cm}
\begin{equation}
\label{equ:M}
M_{i,j} = 
\begin{cases}
PL - G_{i,j} + \Delta_{i,j} L_{i,j} & \text{if } L_{i,j}>0 \\
PL - G_{i,j} & \text{if } L_{i,j} = 0 ,
\end{cases}
\end{equation}
where $PL$ denotes the pathloss value at grid $(i,j)$ calculated by the 3GPP InH model \cite{3gpp_tr38_901_2024}. $G_{i,j}$ denotes antenna gain at grid $(i,j)$ which is obtained from the challenge dataset.

\vspace{-0.3cm}
\subsection{Curriculum learning}
\vspace{-0.2cm}
A curriculum learning strategy is employed to train IPP-Net in three stages. First, we train IPP-Net based on the simulation data for varied indoor scenarios in Task 1. Then, we further generalize IPP-Net based on the simulation data for varied frequency bands in Task 2, starting from the learned network weights from Task 1. Finally, we refine IPP-Net based on the simulation data for varied antenna radiation patterns in Task 3, starting from the learned network weights from Task 2. This staged training process enables IPP-Net to progressively handle more challenging tasks. 

\vspace{-0.3cm}
\section{Experiments}
\label{sec:results}
\vspace{-0.3cm}

The experiments are performed on an NVIDIA RTX3080Ti GPU under the PyTorch framework. For all three tasks, we split the challenge dataset \cite{c0ec-cw74-24} into training and validation sets with a ratio of 9:1. The experiment settings are shown in Table \ref{settings}. 
\vspace{-0.4cm}
\begin{table}[h]
\small
    \centering
    \caption{Experiment settings}
    \vspace{-0.3cm}
    \begin{tabular}{cc}
    \hline
        \textbf{Hyperparameter} & \textbf{Value} \\ \hline
        Learning rate & $10^{-3} \sim 3.125\times10^{-5}$ \\
        ReduceLROnPlateau & factor=0.5, patience=5 \\
        Batch size & 8 \\
        Optimizer & AdamW, weight decay=$10^{-2}$ \\
        Maximum of epochs & 120 \\
        Loss function & RMSE \\ \hline
    \end{tabular}
    \label{settings}
\end{table}
\vspace{-0.4cm}

\begin{figure}[!t]
\centering
\subfloat[Scenario]{
    \begin{minipage}[b]{0.3\linewidth}
    \centering
    \label{fig:1a}
    \includegraphics[scale=0.28]{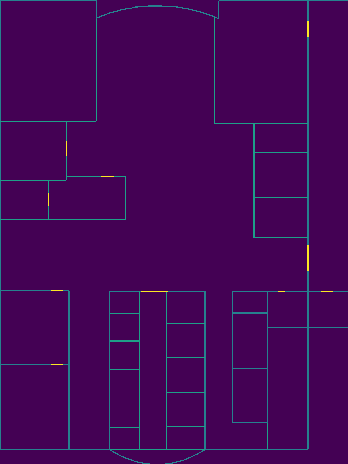}
    \end{minipage}
    }
\subfloat[Prediction]{
    \begin{minipage}[b]{0.3\linewidth}
    \centering
    \label{fig:1b}
    \includegraphics[scale=0.28]{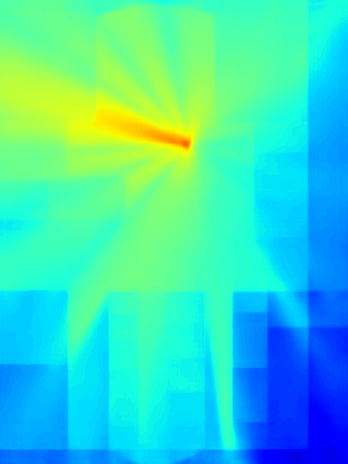}
    \end{minipage}
    }
\subfloat[Target]{
    \begin{minipage}[b]{0.3\linewidth}
    \centering
    \label{fig:1c}
    \includegraphics[scale=0.28]{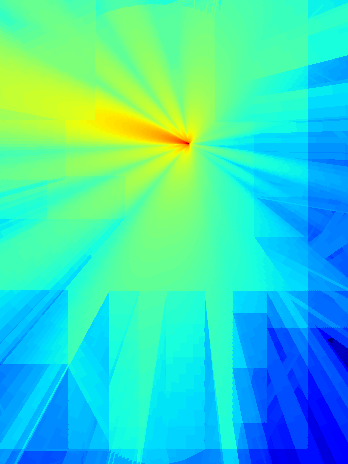}
    \end{minipage}
    }
\vspace{-0.3cm}
\caption{An example radio map predicted by IPP-Net}
\vspace{-0.6cm}
\label{fig}
\end{figure}

The training on Tasks 1-3 takes approximately 20 minutes, four hours, and eight hours, respectively. The training RMSE and the validation RMSE for each training stage are shown in Table \ref{results1}. After the three-stage curriculum learning, the network weights that yield the lowest RMSE on the validation set are saved for the final evaluation. During evaluation, IPP-Net spends an average run-time of 10.51 ms in predicting one radio map, with 0.71 ms on $PL$ calculation, 0.62 ms on $G$ calculation, 5.45 ms for a single pass through IPP-Net, and 3.73 ms for other signal processing components. An example radio map predicted by IPP-Net is shown in Fig. \ref{fig}.

IPP-Net is finally evaluated by the challenge organizing team on an unseen test set of 6 indoor scenarios, 2 frequency bands, and 3 antenna radiation patterns. The RMSE of IPP-Net in the final evaluation is shown in Table \ref{results2}.

\vspace{-0.1cm}
\begin{table}[!ht]
\small
    \centering
    \caption{RMSE in different training stages (dB)}
    \vspace{-0.3cm}
    \tabcolsep=0.45cm
    \begin{tabular}{cccc}
    \hline
        \textbf{} & \textbf{Task 1} & \textbf{Task 2} & \textbf{Task 3} \\ \hline
        Training & 2.8292 & 4.1147 & 3.6593 \\ \hline
        Validation & 4.1093 & 5.2002 & 4.2860 \\ \hline
    \end{tabular}
    \label{results1}
\end{table}

\vspace{-0.4cm}
\begin{table}[!ht]
\small
    \centering
    \caption{RMSE in the final evaluation (dB)}
    \vspace{-0.3cm}
    \begin{tabular}{ccccc}
    \hline
        \textbf{} & \textbf{Task 1} & \textbf{Task 2} & \textbf{Task 3} & \textbf{Final (Weighted)} \\ \hline
        Test & 7.84 & 10.15 & 10.26 & 9.501 \\ \hline
    \end{tabular}
    \label{results2}
\end{table}

\vspace{-0.7cm}
\section{Conclusion}
\vspace{-0.3cm}
A generalizable DNN model for indoor pathloss RMP (IPP-Net) has been proposed in this paper. IPP-Net has achieved a weighted RMSE of 9.501 dB on the three competition tasks and obtained the second overall ranking in \textit{the First Indoor Pathloss Radio Map Prediction Challenge in ICASSP 2025}.

\end{document}